\documentclass[pra,twocolumn,superscriptaddress,showpacs,nofootinbibfloatfix,amsmath,amsfonts,amssymb]{revtex4-1}%
\usepackage{amsmath,amsfonts,amssymb,color}
\usepackage{amsthm}
\usepackage{leftidx}
\usepackage{graphicx}
\usepackage{xcolor}
\usepackage{dcolumn}
\usepackage{bm}
\usepackage{epstopdf}
\usepackage{epsfig}
\usepackage{environ}
\usepackage{pdfcomment}

\usepackage{float}
\usepackage[T1]{fontenc}
\usepackage[latin9]{inputenc}
\usepackage{setspace}
\usepackage{esint}

\begin{document}
	
\preprint{APS/123-QED}

\title{Dimerization induced mobility edges and multiple reentrant localization transitions in non-Hermitian quasicrystals}

\author{Wenqian Han}
\affiliation{%
	College of Physics and Optoelectronic Engineering, Ocean University of China, Qingdao 266100, China
}

\author{Longwen Zhou}
\email{zhoulw13@u.nus.edu}
\affiliation{%
	College of Physics and Optoelectronic Engineering, Ocean University of China, Qingdao 266100, China
}

\date{\today}

\begin{abstract}
Non-Hermitian effects could create rich dynamical and
topological phase structures. In this work, we show that the collaboration
between lattice dimerization and non-Hermiticity could generally bring
about mobility edges and multiple localization transitions
in one-dimensional quasicrystals. Non-Hermitian extensions of the
Aubry-Andr\'e-Harper~(AAH) model with staggered onsite potential and dimerized
hopping amplitudes are introduced to demonstrate our results. 
Reentrant localization transitions due to the interplay between quasiperiodic gain/loss
and lattice dimerization are found. Quantized
winding numbers are further adopted as topological invariants to
characterize transitions among phases with distinct spectrum and transport
nature. Our study thus enriches the family of non-Hermitian quasicrystals
by incorporating effects of lattice dimerization, and offering a convenient
way to modulate localization transitions and mobility edges in non-Hermitian
systems.
\end{abstract}

\maketitle

\section{Introduction\label{sec:Int}}
Non-Hermitian systems have attracted increasing attention over the past
decade due to their abundant dynamical, topological and transport
properties~(see Refs.~\cite{NHRev3,NHRev5,NHRev4,NHRev1,NHRev2,NHRev6} for reviews).
Theoretical studies have uncovered
physics and phenomena that are unique to non-Hermitian settings, such
as the exceptional points~\cite{EP1,EP2,EP3}, non-Hermitian skin effects~\cite{NHSE1,NHSE2,NHSE3,NHSE4}, anomalous localization transitions~\cite{HNM1,HNM2,HNM3,HNM4} and
enlarged symmetry classifications of topological matter~\cite{NHClass1,NHClass2,NHClass3,NHClass4,NHClass5}.
Experimental progress has also made it possible to observe novel
nonunitary dynamics and non-Hermitian topological phases in a
variety of platforms, ranging from cold atoms~\cite{NHExp1,NHExp2,NHExp3}, photonics~\cite{NHExp4,NHExp5,NHExp6,NHExp7}, acoustics~\cite{NHExp11,NHExp12,NHExp13}, electrical
circuits~\cite{NHExp8,NHExp9,NHExp10} to NV center in diamonds~\cite{NHExp14}. These developments further suggest
potential applications of non-Hermitian effects in topological lasers~\cite{TILZ1,TILZ2,TILZ3} and
high-performance sensors~\cite{NHSens1,NHSens2,NHSens3,NHSens4}.

Recently, the interplay between non-Hermitian effects and spatial
quasiperiodicity has been found to generate a rich class of matter
called non-Hermitian quasicrystal (NHQC)~\cite{NHQC0,NHQC1,NHQC2,NHQC3,NHQC4,NHQC5,NHQC6,NHQC7,NHQC8,NHQC9,NHQC10,NHQC11,NHQC12,NHQC13,NHQC14,NHQC15,NHQC16,NHQC17,NHQC18,NHQC19,NHQC20,NHQC201,NHQC21,NHQC22,NHQC23,NHQC24,NHQC25,NHQC26,NHQC27}. In an NHQC, onsite gain
and loss or nonreciprocal hoppings could induce ${\cal PT}$-breaking
transitions of the spectrum from real to complex and localization
transitions of eigenstates between extended and localized phases.
Moreover, in certain situations, a critical phase could appear in
between the extended and localized phases, which holds
energy-dependent mobility edges separating delocalized and localized
states~\cite{NHQC3,NHQC6}. In a Hermitian quasicrystal, critical mobility edge
phases could generally appear when the system possesses long-range
hoppings~\cite{LRQC1,LRQC2,LRQC3} or dimerized lattice structures~\cite{DimerQC1,DimerQC2,DimerQC3}.
However, much less is known about generic ways of inducing and controlling mobility edges and localization transitions in non-Hermitian systems~\cite{NHQC20,NHQC201}.

In this work, we introduce a simple scheme to generate critical phases
with mobility edges and create multiple reentrant localization transitions in
NHQCs by adding experimentally implementable spatial dimerization
to the onsite potential or hopping amplitudes. In Sec.~\ref{sec:The},
 we introduce general theoretical tools that can be
used to unveil the spectral, localization, topological transitions
and mobility edges in NHQCs. In Sec.~\ref{sec:Mod}, we present
models that will be considered in this work and discuss their common
features. In Sec.~\ref{sec:Res}, we reveal how staggered onsite potentials
or dimerized hopping amplitudes could produce critical regions separating
extended and localized phases in NHQCs, and how the change of dimerization
effects could induce reentrant topological transitions among NHQC phases with
distinct spectral and transport nature. In Sec.~\ref{sec:Sum}, we
summarize our results and discuss potential future studies.

\section{Theory\label{sec:The}}
We first outline the theoretical framework that will be
adopted to investigate NHQCs in this work. We focus on one-dimensional (1D) systems with the following
form of lattice Hamiltonian
\begin{equation}
H=\sum_{n\in\mathbb{Z}}(J_{n}^{R}|n\rangle\langle n+1|+J_{n}^{L}|n+1\rangle\langle n|+V_{n}|n\rangle\langle n|).\label{eq:H}
\end{equation}
Here $J_{n}^{R}$ ($J_{n}^{L}$) describes the hopping amplitude from
the $(n+1)$'s ($n$'s) to the $n$'s {[}$(n+1)$'s{]} site of the
lattice. $V_{n}$ is the amplitude of onsite potential. For a
system with periodic boundary condition (PBC), we take the lattice
site index $n=1,...,L$ and identify the basis $|n\rangle=|n+L\rangle$.
The Hamiltonian $H$ can be made non-Hermitian by either setting $J_{n}^{R}\neq(J_{n}^{L})^{*}$
(asymmetric hopping) or $V_{n}\neq V_{n}^{*}$ (onsite gain and loss).
$H$ further describes a quasicrystal if we set $J_{n}^{R,L}$ or
$V_{n}$ to a quasiperiodic function of $n$. In the lattice representation,
the spectrum and states of $H$ can be obtained by solving the eigenvalue
equation $H|\psi\rangle=E|\psi\rangle$, yielding
\begin{equation}
J_{n}^{R}\psi_{n+1}+J_{n-1}^{L}\psi_{n-1}+V_{n}\psi_{n}=E\psi_{n}.\label{eq:Seq}
\end{equation}
Here $\psi_{n}$ represents the amplitude of wave function $|\psi\rangle=\sum_{n}\psi_{n}|n\rangle$
on the $n$th lattice site. For a lattice of length $L$, there are
$L$ eigenenergies $E=\{E_{j}|j=1,...,L\}$, whose values
can in general be complex if $H\neq H^{\dagger}$.

In a system described by $H$, the presence of non-Hermiticity may
induce three types of transitions~\cite{NHQC1}. On the spectral side, if $H$ possesses
the ${\cal PT}$ or other types of symmetry that could make it pseudo-Hermitian,
its spectrum can be real and the growth of its non-Hermitian parameters
may induce a real-to-complex (e.g., the ${\cal PT}$-breaking) transition in the spectrum. Such a transition
can be identified by looking at the maximum of the imaginary parts
of $E$ over all eigenstates, i.e., 
\begin{equation}
\max({\rm Im}E)=\max_{j\in\{1,...,L\}}(|{\rm Im}E_{j}|).\label{eq:MaxImE}
\end{equation}
It is clear that the value of $\max({\rm Im}E)$ grows from zero
to finite when the spectrum of $H$ changes from real to complex, and
vice versa. Besides, not all eigenstates may take complex energies
after the spectrum transition. To reveal this fact, we introduce
the density of states (DOSs) with complex eigenenergies
\begin{equation}
\rho=N({\rm Im}E\neq0)/L,\label{eq:DOS}
\end{equation}
where $N({\rm Im}E\neq0)$ denotes the number of states whose energies
have finite imaginary parts. In the deep non-Hermitian region we
would expect $\rho\rightarrow1$.

An NHQC could generally possess a localized phase,
an extended phase, and a critical phase in which localized and extended
states coexist and are separated by a mobility edge~\cite{NHQC3,NHQC6}. Different characteristics
of these phases can be extracted from the level-spacing statistics and
inverse/normalized participation ratios of the states. Assuming
the eigenenergies $\{E_{j}|j=1,...,L\}$ are sorted
by their real parts and denoting the real gap between the $j$th and $(j-1)$th
energy levels as $\varepsilon_{j}={\rm Re}E_{j}-{\rm Re}E_{j-1}$,
we define adjacent gap ratios (AGRs) as $g_{j}=\min(\varepsilon_{j},\varepsilon_{j+1})/\max(\varepsilon_{j},\varepsilon_{j+1})$
for $j=2,...,L-1$. Here $\min(a,b)$ and $\max(a,b)$ yield the minimum
and maximum of $a$ and $b$, respectively. The statistical feature
of level-spacing is encoded in the average of AGRs over all states,
which is defined as
\begin{equation}
\overline{g}=\frac{1}{L}\sum_{j}g_{j}.\label{eq:AveAGR}
\end{equation}
In the limit $L\rightarrow\infty$, we have $\overline{g}\rightarrow0$
if all bulk states are extended, and $\overline{g}$ approaches a
constant in the localized phase~\cite{NHRMT1,NHRMT2,NHRMT3,NHRMT4,NHRMT5}. In the critical phase, the
value of $\overline{g}$ is non-universal and changes
between zero and its upper bound obtained for the localized phase.
Meanwhile, if $|\psi_{j}\rangle=\sum_{n=1}^{L}\psi_{n}^{j}|n\rangle$
is a normalized eigenstate of $H$ with energy $E_{j}$, we define
its inverse participation ratio (IPR) and normalized participation
ratio (NPR) as ${\rm IPR}_{j}=\sum_{n=1}^{L}|\psi_{n}^{j}|^{4}$ and
${\rm NPR}_{j}=(L\sum_{n=1}^{L}|\psi_{n}^{j}|^{4})^{-1}$.
In the thermodynamic limit $L\rightarrow\infty$, we have ${\rm IPR}_{j}\rightarrow0$
(${\rm IPR}_{j}\rightarrow\xi_{j}^{-1}$) and ${\rm NPR}_{j}\rightarrow1$
(${\rm NPR}_{j}\rightarrow0$) if $|\psi_{j}\rangle$ is an extended
(a localized) state, where the localization length $\xi_{j}$ could
be energy-dependent~\cite{CMTBook1}. With the help of IPRs and NPRs, we can distinguish
the three phases and capture the transitions among them from the
following quantities:
\begin{equation}
\max({\rm IPR})=\max_{j\in\{1,...,L\}}({\rm IPR}_{j}),\label{eq:IPRmax}
\end{equation}
\begin{equation}
\min({\rm IPR})=\min_{j\in\{1,...,L\}}({\rm IPR}_{j}),\label{eq:IPRmin}
\end{equation}
\begin{equation}
\eta=\log_{10}(\langle{\rm IPR}\rangle\langle{\rm NPR}\rangle).\label{eq:ETA}
\end{equation}
Here $\langle{\rm IPR}\rangle=\frac{1}{L}\sum_{j=1}^{L}{\rm IPR}_{j}$ and
$\langle{\rm NPR}\rangle=\frac{1}{L}\sum_{j=1}^{L}{\rm NPR}_{j}$ denote the
average of IPRs and NPRs over all states. When $L\rightarrow\infty$,
we would have $\max({\rm IPR})\rightarrow0$ if the system resides
in an extended phase, $\min({\rm IPR})>0$ if the system stays in a localized
phase, and $\eta$ to be finite if the system is in a critical phase with mobility
edges~\cite{DimerQC2}. The AGRs, IPRs and NPRs
can thus be used to distinguish extended, localized and critical
phases of an NHQC from complementary perspectives.
Note in passing that in this work, our calculations 
are performed under the PBC. 
In this case, the non-Hermitian skin effect is absent and will
not affect the spatial distribution of bulk states. 
Therefore, we compute the IPR and NPR for
all bulk states using right eigenvectors, whereas using left
eigenvectors will generate equivalent results concerning Eqs.~(\ref{eq:IPRmax})-(\ref{eq:ETA}).

When the energies of an NHQC possess imaginary parts,
they may develop loop structures on the complex plane.
Interestingly, the emergence of these complex energy loops was found
to be related to the localization transitions in NHQCs.
One can then introduce spectral winding numbers as
topological order parameters to characterize phases with different
transport nature and signify the transitions among them in an NHQC~\cite{NHQC1,NHQC5}.
The generic definition of such a winding number is
\begin{equation}
w_{\ell}=\lim_{L\rightarrow\infty}\int_{0}^{2\pi}\frac{d\theta}{2\pi i}\partial_{\theta}\ln\{\det[H(\theta)-{\cal E}_{\ell}]\}.\label{eq:W}
\end{equation}
Here $\theta$ can be viewed as a flux penetrating through the ring
formed by the system under PBC. 
The model Hamiltonian $H$ in Eq.~(\ref{eq:H}) can be changed to its $\theta$-dependent form
$H(\theta)$ by either setting $J_n^{R(L)}$ to $J_n^{R(L)}e^{+(-)i\theta/N}$ or adding a phase shift $\theta/N$ to $V_n$, where
$N$ is the number of cells of the lattice. For the models considered in this work, the ways of introducing phase factor $\theta$
to $H$ are summarized in Table \ref{tab:Mod}.
${\cal E}_{\ell}$
is a model-dependent base energy on the complex plane, and $w_{\ell}$
counts the number of times the spectrum of $H(\theta)$ winds around
${\cal E}_{\ell}$ when $\theta$ is swept over a period. If the system
has no critical phases, there is only one ${\cal E}_{\ell}={\cal E}_{1}$,
which can be chosen as the real part of energy of the first eigenstate
of $H$ whose IPR deviates from zero. The transition of the system
from the extended to localized phases would then accompany the quantized
jump of $w_{1}$. If the extended and localized phases of the system
are further separated by a critical phase, another base energy ${\cal E}_{\ell}={\cal E}_{2}$
is needed, which can be chosen as the real part of energy of the last
eigenstate of $H$ whose IPR departures from zero. In this case, we
expect $w_{1}$ ($w_{2}$) to take a quantized jump when the system
goes from the extended (critical) phase to the critical (localized)
phase~\cite{NHQC20}. Within a given phase, the winding number will
be pinned to a quantized value and can thus be interpreted as a
topological order parameter of the corresponding NHQC phase. 

We are now ready to introduce explicit models with
staggered onsite potentials or dimerized hopping amplitudes, and show
how the interplay between lattice dimerization and non-Hermitian effects could
create rich phase and transition patterns in NHQCs. 

\section{Model\label{sec:Mod}}
\begin{table*}
	\begin{centering}
		\begin{tabular}{|c|c|c|c|}
			\hline 
			Model index & Hopping amplitudes $J_{n}^{L,R}$ & Onsite potential $V_n$ & Onsite potential with phase shift $V_n(\theta)$\tabularnewline
			\hline 
			\hline 
			M1 & $J$ & $Ve^{i2\pi\alpha n}+(-1)^{n}\Delta$ & $Ve^{i(2\pi\alpha n+\theta/N)}+(-1)^{n}\Delta$\tabularnewline
			\hline 
			M2 & $J+(-1)^{n}\Lambda$ & $Ve^{i2\pi\alpha n}$ & $Ve^{i(2\pi\alpha n+\theta/N)}$\tabularnewline
			\hline 
			M3 & $J$ & $V\cos(2\pi\alpha n+i\gamma)+(-1)^{n}\Delta$ & $V\cos(2\pi\alpha n+i\gamma+\theta/N)+(-1)^{n}\Delta$\tabularnewline
			\hline 
		\end{tabular}
		\par\end{centering}
	\caption{NHQCs with staggered onsite potential
		or dimerized hopping amplitudes. The system parameters $J,V,\Delta,\Lambda,\gamma\in\mathbb{R}$,
		and the lattice site index $n\in\mathbb{Z}$. $\alpha$ is irrational and set as $\frac{\sqrt{5}-1}{2}$ throughout this work. $N=L/2$ is the number of dimerized cells in the lattice of length $L$. In the calculation of winding numbers, the Hamiltonians $H$ of M1--M3 are replaced by $H(\theta)$ via changing their $V_n$ in the third column to the corresponding $V_n(\theta)$ in the last column.\label{tab:Mod}}
\end{table*}
To uncover the impact of lattice dimerization on NHQCs,
we start with the model introduced in Ref.~\cite{NHQC2}, whose
Hamiltonian takes the form $H_{0}=\sum_{n\in\mathbb{Z}}(J|n\rangle\langle n+1|+{\rm H.c.}+Ve^{i2\pi\alpha n}|n\rangle\langle n|)$.
Here $J,V\in\mathbb{R}$ and $\alpha=\frac{\sqrt{5}-1}{2}$. As a
non-Hermitian extension of the AAH model~\cite{AAH1,AAH2,AAH3}, $H_{0}$ forms a minimal
construction of an NHQC. It was proved that when
$|V|<|J|$, the spectrum of $H_{0}$ is real and all its eigenstates
are extended. When $|V|>|J|$, the spectrum becomes complex
and all eigenstates are localized. Therefore, the system described
by $H_{0}$ undergoes a ${\cal PT}$-breaking transition
together with a localization transition at $|V|=|J|$, which can be
topologically characterized by the quantized jump of a spectral winding
number~\cite{NHQC24}. Note that no critical phases and mobility edges are found
in $H_{0}$, and all states undergo the same
localization transition when $|V|$ switches from
below to above $|J|$, with the common Lyapunov exponent $\lambda=\ln|V/J|$~\cite{NHQC2}.

Another NHQC model that will be employed has the Hamiltonian $H_1=\sum_{n\in\mathbb{Z}}(J|n\rangle\langle n+1|+{\rm H.c.}+V\cos(2\pi\alpha n+i\gamma)|n\rangle\langle n|)$, where $i\gamma$ introduces an imaginary phase
shift in the superlattice potential. It was found that when $\gamma=\gamma_c=\ln|2J/V|$,
the states of $H_1$ could undergo a transition from an extended phase with real spectrum ($|\gamma|<|\gamma_c|$) to a localized phase with complex spectrum ($|\gamma|>|\gamma_c|$)~\cite{NHQC1}.
This transition is further accompanied by the quantized jump of a spectral winding number $w$ for zero to $-1$~\cite{NHQC1}.
However, there are also no signatures of critical mobility edge phases in the system described by $H_1$, and all eigenstates experience the same spectral and localization transitions at $\gamma=\gamma_c$.

In this work, we use dimerized versions of $H_{0}$ and $H_1$ to showcase
our main results. We consider three different extensions,
which are denoted by M1--M3. Their Hamiltonians share the form of
Eq.~(\ref{eq:H}) with components listed in Table \ref{tab:Mod}.
In M1--M3, the strength of hopping dimerization and staggered
onsite potential are separately controlled by the parameters $\Lambda$
and $\Delta$. For the calculations presented below,
we choose $J=1$ as the unit of energy, let $\alpha=(\sqrt{5}-1)/2$
be the inverse golden ratio and assuming PBC
throughout. It will be shown that the dimerization effects could generically
induce critical phases with mobility edges, and furthermore trigger
alternated and reentrant localization transitions in NHQCs. 

\section{Results\label{sec:Res}}

We now investigate the spectral, localization and topological
transitions in M1--M3 with the tools introduced in Sec.~\ref{sec:The}.
We first consider the effect of lattice dimerization
in onsite potential (M1) and hopping amplitudes
(M2) in Subsecs.~\ref{subsec:M1} and \ref{subsec:M2}.
Both of these two types of dimerization are found to induce
critical phases with mobility edges and multiple localization transitions
in the presence of non-Hermitian quasiperiodic potential. In Subsec.~\ref{subsec:M3}, 
we unveil intriguing patterns of reentrant localization transitions in M3, which are due to the interplay between onsite dimerization and dissipation.

\subsection{M1: Effect of staggered onsite potential\label{subsec:M1}}

\begin{figure}
	\begin{centering}
		\includegraphics[scale=0.47]{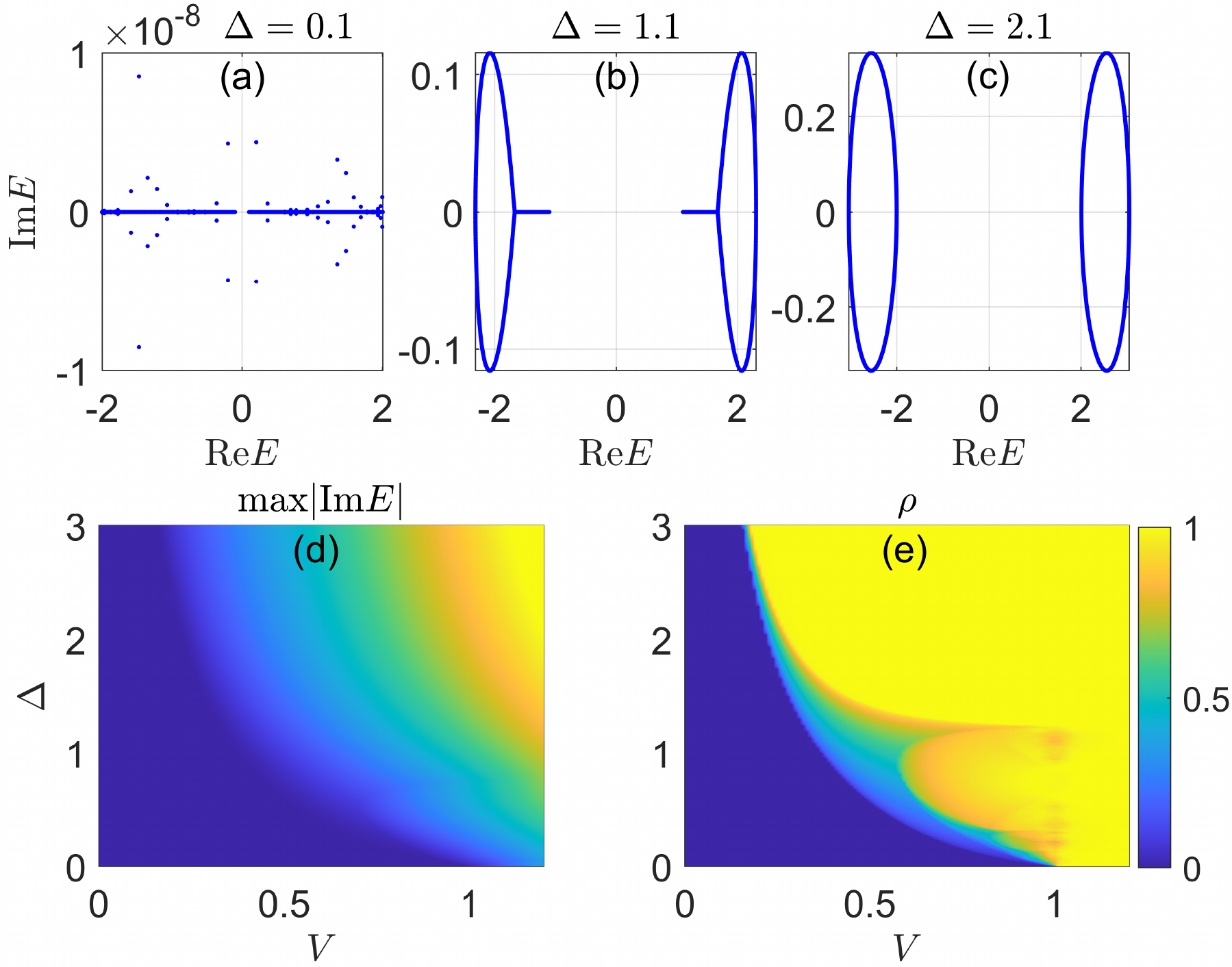}
		\par\end{centering}
	\caption{Spectral properties of M1 under the PBC. The length
		of lattice is $L=2584$ for all panels. (a)--(c) show the spectrum
		of M1 at $V=0.5$ with the increase of staggered onsite potential
		$\Delta$. (d) exhibits the maximum of imaginary parts of
		energies versus $V$ and $\Delta$. (e) denotes the DOSs
		with nonvanishing imaginary parts of energies. (d) and
		(e) share the same color bar.\label{fig:M1E}}
\end{figure}

Referring to Table \ref{tab:Mod}, the M1 corresponds
to a non-Hermitian variant of the AAH model plus a staggered onsite
potential. Following Eq.~(\ref{eq:Seq}), the Hamiltonian of M1 reads
$H=\sum_{n}(|n\rangle\langle n+1|+{\rm H.c.}+V_{n}|n\rangle\langle n|)$
and its eigenvalue equation takes the form $\psi_{n+1}+\psi_{n-1}+V_{n}\psi_{n}=E\psi_{n}$,
where $V_{n}=Ve^{i2\pi\alpha n}+(-1)^{n}\Delta$ and we have set $J=1$
as the unit of energy. Since $V_{n}=V_{-n}^{*}$, M1 possesses the
${\cal PT}$-symmetry, implying that its spectrum could be real in
certain parameter regions. Under the PBC,
the spectrum of M1 for a typical set of onsite dimerization $\Delta$
are shown in Figs.~\ref{fig:M1E}(a)--(c). We observe that with the
increase of $\Delta$, the spectrum could change from purely
real {[}Fig.~\ref{fig:M1E}(a){]} to a mixture of real and complex
energies {[}Fig.~\ref{fig:M1E}(b){]}, and finally be purely
complex {[}Fig.~\ref{fig:M1E}(c){]}. Therefore, the presence of staggered
onsite potential in M1 could not only induce a ${\cal PT}$-breaking
transition of the spectrum, but also create an intermediate region
with coexisting real and complex eigenenergies, which is absent when $\Delta=0$. Compared with the single transition
of $H_{0}$ as mentioned in Sec.~\ref{sec:Mod}, the introduction
of a staggered onsite potential could clearly enrich the spectral
patterns of NHQCs. In Fig.~\ref{fig:M1E}(d),
we further show the maximum of the imaginary part of spectrum versus
the quasiperiodic non-Hermitian potential $V$ and onsite dimerization
$\Delta$. For $V\in(0,1)$, we find a ${\cal PT}$-breaking transition
of the spectrum from real ($\max|{\rm Im}E|=0$) to complex ($\max|{\rm Im}E|>0$)
with the increase of either $V$ or $\Delta$. For $V>1$, the system
enters a phase with only complex energies, and the change of
$\Delta$ would not create further spectral transitions. In Fig.~\ref{fig:M1E}(e),
we present the DOSs {[}Eq.~(\ref{eq:DOS}){]} with complex eigenenergies
$\rho$ versus $V$ and $\Delta$, which show three qualitatively
distinct regions. At the bottom left corner (in blue), we have $\rho=0$
and all eigenstates of $H$ have real energies. At the top right corner
(in yellow), we have $\rho=1$ and the eigenenergies of all states
are complex. In between the former two regions, we find $0<\rho<1$,
implying the eigenstates therein form a mixture of real
and complex energies. Notably, this intermediate region is absent
without the dimerization in onsite potential. Furthermore, as will
be discussed shortly, the states in these three regions possess
distinct localization and topological features.

\begin{figure}
	\begin{centering}
		\includegraphics[scale=0.5]{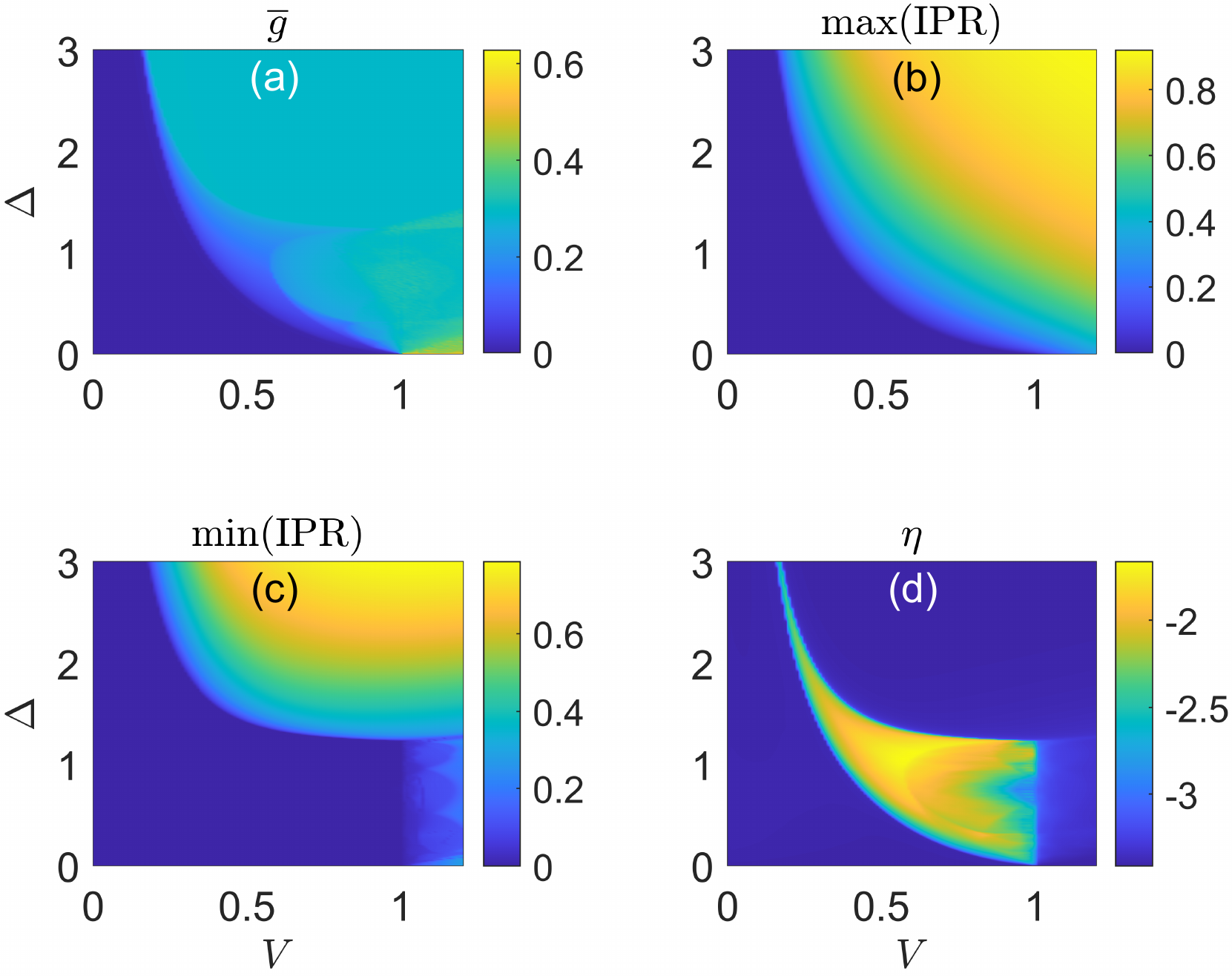}
		\par\end{centering}
	\caption{State properties of M1, computed with the length
		of lattice $L=2584$ under the PBC. (a) shows the averaged AGRs {[}Eq.~(\ref{eq:AveAGR}){]}. (b) and (c) show the maximum
		{[}Eq.~(\ref{eq:IPRmax}){]} and minimum {[}Eq.~(\ref{eq:IPRmin}){]}
		of IPRs over all states at different system parameters $(V,\Delta)$.
		(d) presents the function $\eta$ {[}Eq.~(\ref{eq:ETA}){]}, which
		is finite only in the critical phase with mobility edge.\label{fig:M1IPR}}
\end{figure}

To decode the localization transitions in
M1, we study its averaged AGRs {[}Eq.~(\ref{eq:AveAGR}){]}, IPRs {[}Eqs.~(\ref{eq:IPRmax})--(\ref{eq:IPRmin}){]} 
and the measure $\eta$
{[}Eq.~(\ref{eq:ETA}){]} characterizing the presence of a critical
phase with mobility edge. In Fig.~\ref{fig:M1IPR}(a), we find three
different regions in $\overline{g}$ versus
$V$ and $\Delta$, which have one-to-one
correspondences with the three regions of spectrum in Fig.~\ref{fig:M1E}(e).
Therefore, we expect M1 to show an extended phase ($\overline{g}=0$)
with real spectrum, a localized phase ($\overline{g}\simeq0.3$) with
purely complex spectrum, and a critical phase $0<\overline{g}\lesssim0.3$
in which eigenstates with real and complex energies coexist and are
separated by a mobility edge. These inferences are further confirmed
by the numerical results of $\max({\rm IPR})$ {(}$=0$ only if all
eigenstates are extended{)}, $\min({\rm IPR})$ {(}$>0$ only if all
eigenstates are localized{)} and $\eta$ (finite only if localized
and extended states coexist), as presented in Figs.~\ref{fig:M1IPR}(b)--\ref{fig:M1IPR}(d).
Therefore, the presence of onsite dimerization and its interplay with
the non-Hermitian quasiperiodicity could indeed yield a critical phase
with mobility edge, and induce multiple transitions
in NHQCs. 

\begin{figure}
	\begin{centering}
		\includegraphics[scale=0.48]{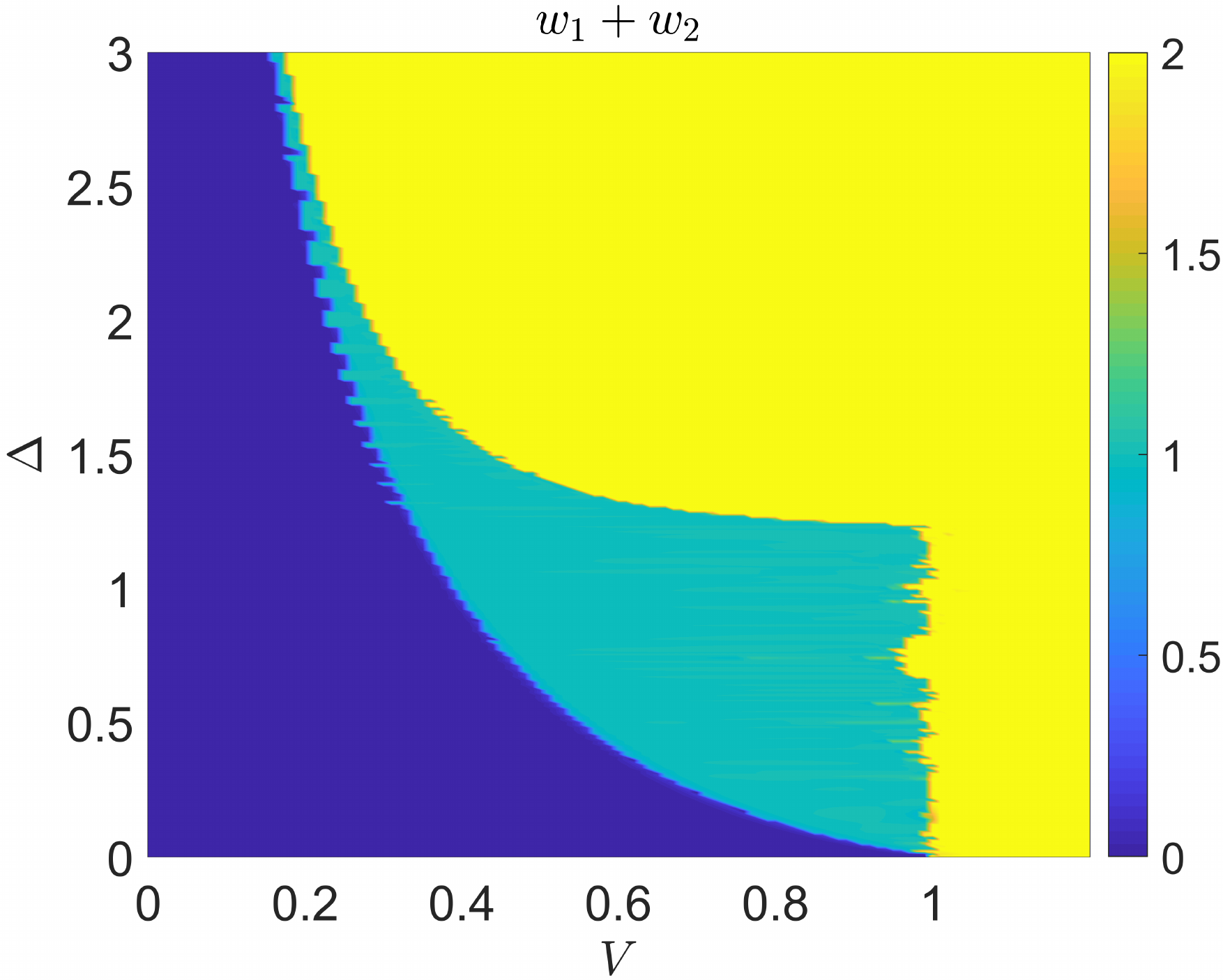}
		\par\end{centering}
	\caption{Topological phase diagram of M1 under the PBC.
		The regions in blue, green and yellow correspond to the extended,
		critical and localized NHQC phases with winding numbers $(w_{1},w_{2})=(0,0)$,
		$(1,0)$ and $(1,1)$, respectively. The length of lattice is set
		as $L=610$ in the calculation.\label{fig:M1WN}}
\end{figure}

The transitions among extended, critical, and localized phases in
M1 can be further attached with quantized jumps of topological invariants.
According to Eq.~(\ref{eq:W}), we construct a pair of winding numbers
$w_{1}$ and $w_{2}$ to characterize the transitions from extended
to critical and from critical to localized phases in M1, respectively.
Note that for M1--M2, the phase shift in Eq.~(\ref{eq:W})
is introduced by setting $Ve^{i2\pi\alpha n}\rightarrow Ve^{i(2\pi\alpha n+2\theta/L)}$
in $V_{n}$, where $L/2$ corresponds to the number of dimerized cells.
By calculating $(w_{1},w_{2})$, we obtain the topological
phase diagram of the system, as presented in Fig.~\ref{fig:M1WN}.
We find a region (in blue) with winding numbers
$(w_{1},w_{2})=(0,0)$, which is in coincidence with the extended phase
with real spectrum, a second region (in green) with $(w_{1},w_{2})=(1,0)$,
which is consistent with the critical phase with partially real
spectrum, and a third region (in yellow) with $(w_{1},w_{2})=(1,1)$
coinciding with the localized phase with purely complex spectrum
in Figs.~\ref{fig:M1E} and \ref{fig:M1IPR}. These results clearly
demonstrate that the spectral winding numbers $(w_{1},w_{2})$ could
be employed as topological order parameters to distinguish the three
NHQC phases of M1 and characterize the transitions among them.
Put together, the existence of staggered onsite potential could
indeed induce ${\cal PT}$-breaking, multiple localization
and topological transitions, and also yield a critical phase with
mobility edge. It can therefore be used as a flexible knob to
tune and even create intriguing phases and transitions in NHQCs.

In Sec.~\ref{subsec:M2}, we show that multiple localization transitions
and mobility edges could also be induced in NHQCs by adding dimerization
to the hopping amplitudes.

\subsection{M2: Effect of dimerized hopping amplitudes\label{subsec:M2}}

We now investigate a non-Hermitian AAH model with
dimerized hopping amplitudes, whose Hamiltonian takes the form
$H=\sum_{n}(J_{n}|n\rangle\langle n+1|+{\rm H.c.}+V_{n}|n\rangle\langle n|)$,
with $J_{n}=1+(-1)^{n}\Lambda$ and $V_{n}=Ve^{i2\pi\alpha n}$ as
given by M2 in Table \ref{tab:Mod}. The resulting eigenvalue equation
is $J_{n}\psi_{n+1}+J_{n-1}\psi_{n-1}+V_{n}\psi_{n}=E\psi_{n}$,
and the uniform part of hopping is set to $J=1$.
In experiments, the dimerized hopping amplitudes can be
realized in various physical platforms. For example, in cold atom systems,
it can be engineered by superimposing two standing optical waves of wavelengths
$\lambda$ and $2\lambda$ to generate a lattice potential of the form 
$U\sin^2(2\pi x/\lambda+\phi/2)+U'\sin^2(4\pi x/\lambda+\pi/2)$, where $x$
denotes the position and $\phi$ represents a controllable phase 
between the two standing wave fields of strengths $U$ and $U'$~\cite{SSH1,SSH2,SSH3}.
Since $J_{n}=J_{-n}^{*}$
and $V_{n}=V_{-n}^{*}$, M2 also obeys the ${\cal PT}$-symmetry,
which means that its spectrum can be real in some parameter domains.
Note that a different non-Hermitian extension of M2 was considered
in Ref.~\cite{NHQC20}, and dimerization-induced intermediate phases with
mobility edges were also observed.

\begin{figure}
	\begin{centering}
		\includegraphics[scale=0.49]{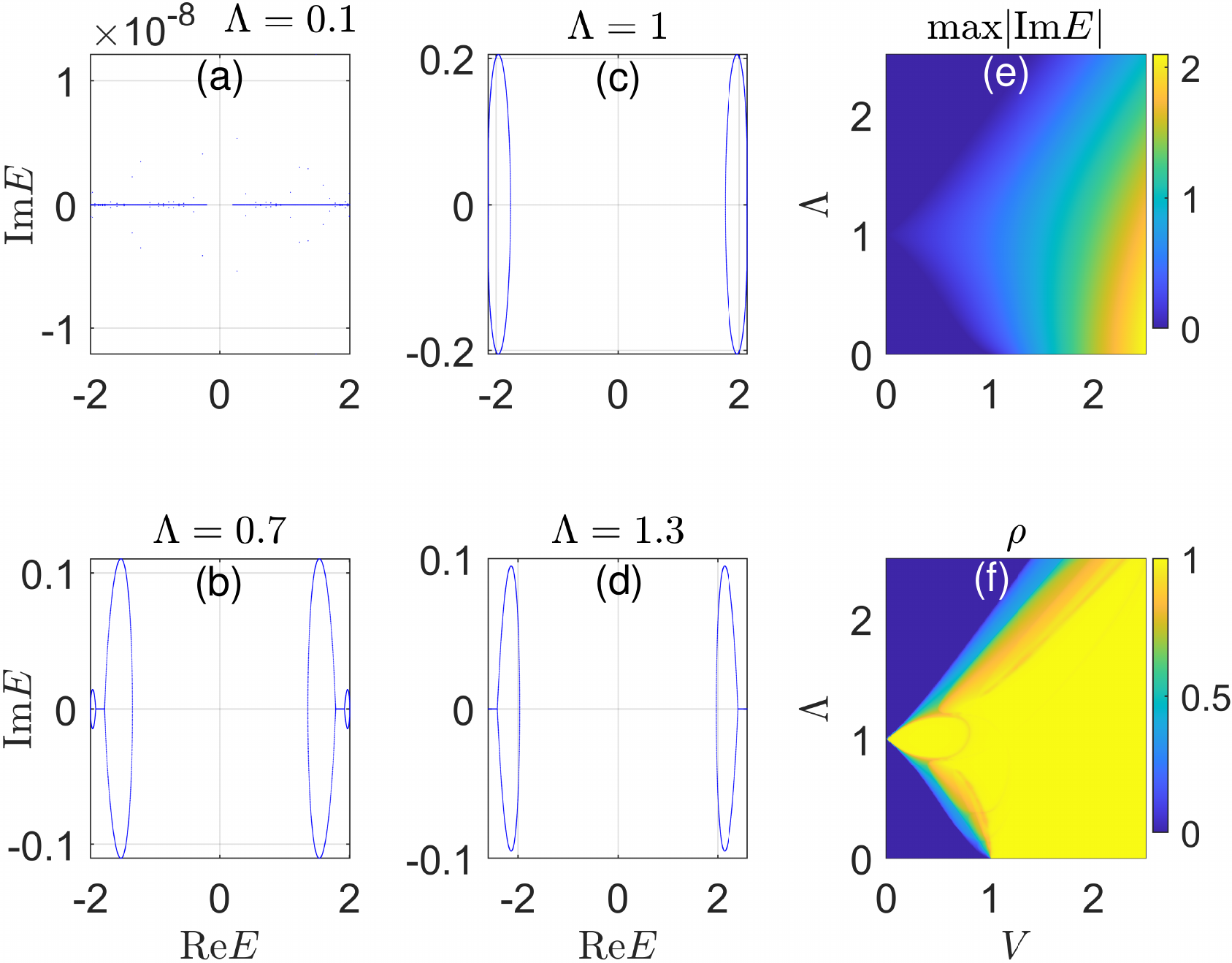}
		\par\end{centering}
	\caption{Spectral properties of M2 under the PBC. The length
		of lattice is $L=2584$ for all panels. (a)--(d) show the spectrum
		at $V=0.5$ with the increase of staggered hopping amplitude
		$\Lambda$. (e) presents the maximum of imaginary parts of energy
		versus $V$ and $\Lambda$. (f) shows the DOSs
		with nonvanishing imaginary parts of energy. \label{fig:M2E}}
\end{figure}

When $V\in(0,1)$, M2 undergoes four spectral transitions
with the increase of hopping dimerization $\Lambda$. Typical spectra
of the system before and after these transitions are shown in Figs.~\ref{fig:M2E}(a)--\ref{fig:M2E}(d).
With the increase of $\Lambda$,
M2 first goes through a ${\cal PT}$-breaking transition from a
real spectrum {[}Fig.~\ref{fig:M2E}(a){]} to a mixed spectrum with
both real and complex eigenvalues {[}Fig.~\ref{fig:M2E}(b){]}.
At a larger $\Lambda$, states with real energies vanish and the
spectral becomes purely complex {[}Fig.~\ref{fig:M2E}(c){]}. However,
with the further increase of $\Lambda$, real eigenvalues in the spectrum
re-emerge and the system goes back to a phase with coexisting
real and complex energies, as shown in Fig.~\ref{fig:M2E}(d). When
$\Lambda$ becomes even larger, M2 could again enter the phase
with real spectrum and the ${\cal PT}$-symmetry is recovered, which
corresponds to the region with $\max|{\rm Im}E|=0$ at the top left
corner of Fig.~\ref{fig:M2E}(e). The DOSs with complex energies
shown in Fig.~\ref{fig:M2E}(f) further confirms the observed alternating
transitions among real, mixed and complex spectrum regions. The physical
mechanism behind these reentrant spectral transitions may be understood
as follows. With the increase of $\Lambda$, the averaged hopping
amplitudes among lattice sites, which may be expressed as a function
$f(J-\Lambda,J+\Lambda)$ does not change monotonically with $\Lambda$.
Since the spectrum transitions in M2 are originated from the competition
between hopping and onsite energy scales, the non-monotonic
behavior of $f(J-\Lambda,J+\Lambda)$ in $\Lambda$ implies the non-monotonic
process of spectral transitions.

\begin{figure}
	\begin{centering}
		\includegraphics[scale=0.5]{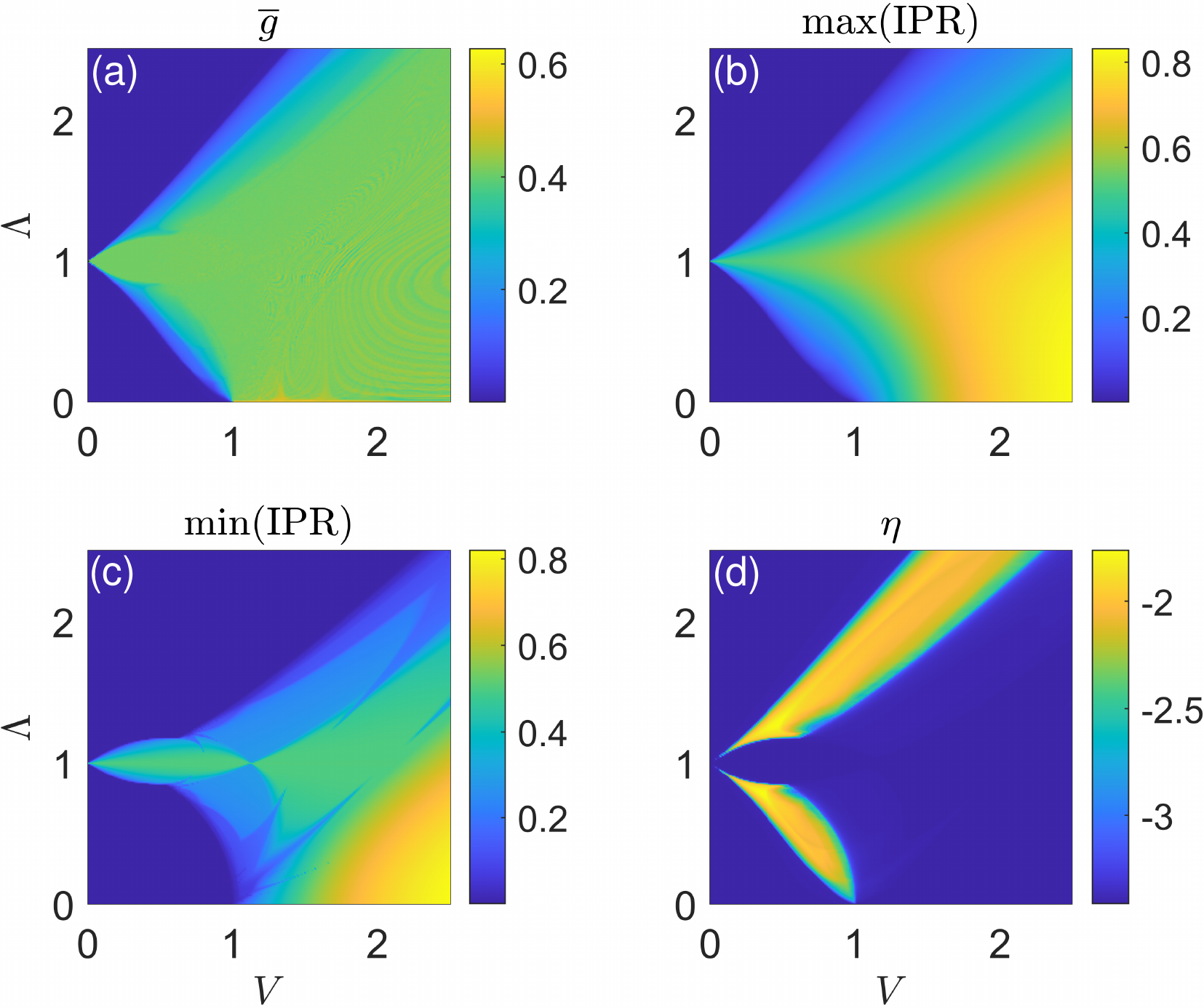}
		\par\end{centering}
	\caption{State properties of M2, computed with the length
		of lattice $L=2584$ under the PBC. (a) shows the average of AGRs over
		all states {[}Eq.~(\ref{eq:AveAGR}){]}. (b) and (c) show the maximum
		{[}Eq.~(\ref{eq:IPRmax}){]} and minimum {[}Eq.~(\ref{eq:IPRmin}){]}
		of IPRs at different system parameters $(V,\Lambda)$.
		(d) shows the quantity $\eta$ {[}Eq.~(\ref{eq:ETA}){]}, whose value
		is finite only in critical phases with mobility edges.\label{fig:M2IPR}}
\end{figure}

Similar to the cases encountered in M1, the spectral transitions
in M2 are also accompanied by state transitions among extended, critical,
and localized phases. This is demonstrated by the results presented
in Fig.~\ref{fig:M2IPR}, which describe the averaged AGRs {[}Fig.~\ref{fig:M2IPR}(a){]}, 
IPRs {[}Fig.~\ref{fig:M2IPR}(b)--(c){]}
and the measure of critical phase with mobility edges $\eta$ {[}Fig.~\ref{fig:M2IPR}(d){]}. 
From the minimum and maximum of IPRs, we can
clearly see two distinct phase boundaries at $\max({\rm IPR})=0\rightarrow>0$
and $\min({\rm IPR})=0\rightarrow>0$, respectively, across which
the system changes from the extended to critical and from the critical
to localized phases. These phase boundaries coincide with the
parameters at which the DOSs with complex energies change
from $0\rightarrow\rho\in(0,1)$ and from $\rho\in(0,1)\rightarrow1$ in Fig.~\ref{fig:M2E}(f).
The critical phases in which extended and localized states coexist
are highlighted by the regions with finite values of $\eta$ in Fig.~\ref{fig:M2IPR}(d). 
These two sets of states are separated in energy by mobility edges.
Notably, there are two such
critical regions separated by a localized phase for $V\in(0,1)$.
These phases are originated from the interplay between dimerized hoppings
and non-Hermitian quasiperiodic potential. Therefore, with the increase
of $\Lambda$, the system could undergo multiple
and reentrant transitions among extended, critical, and localized
NHQC phases. Each transition is accompanied by a drastic change
in the structure of spectrum.

\begin{figure}
	\begin{centering}
		\includegraphics[scale=0.47]{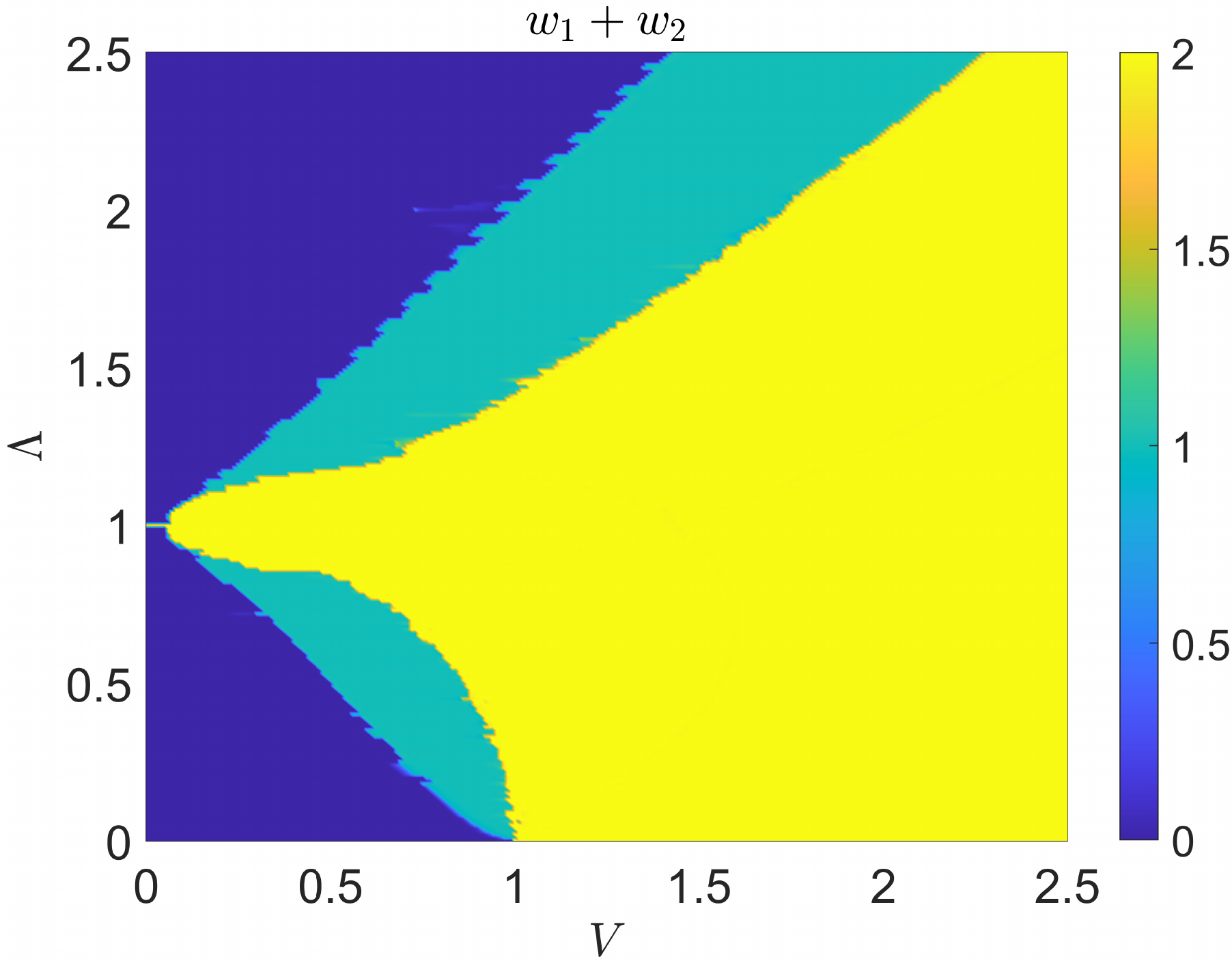}
		\par\end{centering}
	\caption{Topological phase diagram of M2 under the PBC.
		Regions in blue, green and yellow correspond to the extended,
		critical and localized NHQC phases with spectral winding numbers $(w_{1},w_{2})=(0,0)$,
		$(1,0)$, and $(1,1)$, respectively. The length of lattice is chosen to be
		$L=610$. \label{fig:M2WN}}
\end{figure}

With the help of Eq.~(\ref{eq:W}), we obtain the spectral winding
numbers $(w_{1},w_{2})$ of M2, leading to the topological phase
diagram presented in Fig.~\ref{fig:M2WN}. Together with the results shown in
Figs.~\ref{fig:M2E} and \ref{fig:M2IPR}, we see that there are indeed
three distinct topological NHQC phases in M2. The extended phase (with
real spectrum) and localized phase (with complex spectrum) possess
winding numbers $(w_{1},w_{2})=(0,0)$ and $(w_{1},w_{2})=(1,1)$,
whereas the critical phase (with mobility edge) has $(w_{1},w_{2})=(1,0)$.
When $V\in(0,1)$, we could encounter four topological transitions
with the increase of hopping dimerization $\Lambda$. At a fixed
$\Lambda$ ($\neq0,1$), M2 will first change from the extended
to critical phase, and finally enter the localized phase with
the increase of non-Hermitian potential $V$. The critical
phases with $(w_{1},w_{2})=(1,0)$ separating the extended and localized
ones are present only if $\Delta\neq0$ in
M2. Put together, we conclude that both dimerized onsite and
hopping modulations could induce critical regions with mobility edges in NHQCs and control
transitions among phases with distinct spectral, localization and
topological nature.

\subsection{M3: Multiple reentrant localization transitions\label{subsec:M3}}

\begin{figure}
	\begin{centering}
		\includegraphics[scale=0.48]{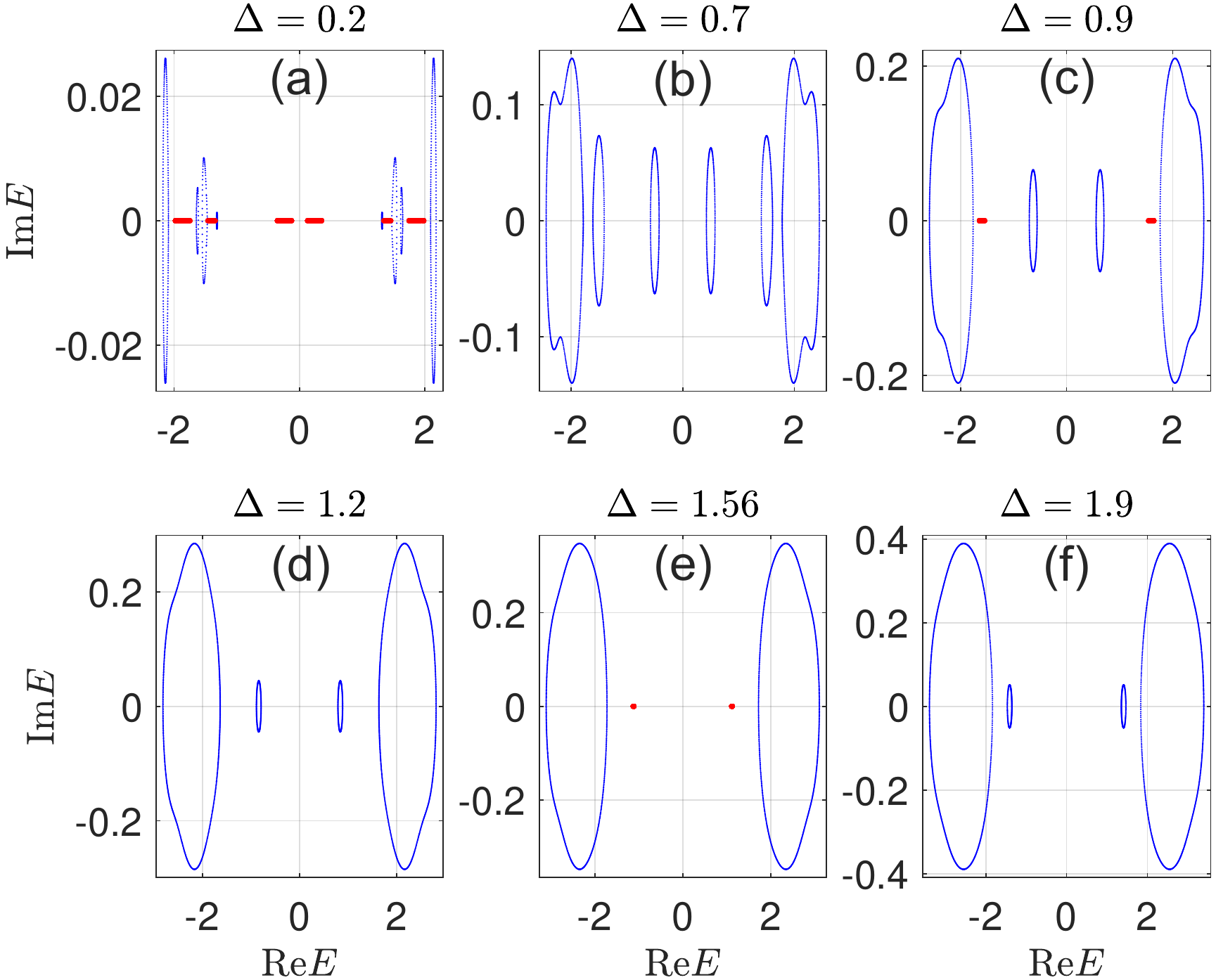}
		\par\end{centering}
	\caption{Typical spectra of M3, obtained for
		a lattice of length $L=2584$ under the PBC. States with real (complex) eigenenergies
		are denoted by red stars (blue dots). The values of dimerization parameter $\Delta$
		are listed in the captions of (a)--(f). Other system parameters
		are set as $(J,V,\gamma)=(1,1,0.5)$.\label{fig:M3E}}
\end{figure}

\begin{figure}
	\begin{centering}
		\includegraphics[scale=0.47]{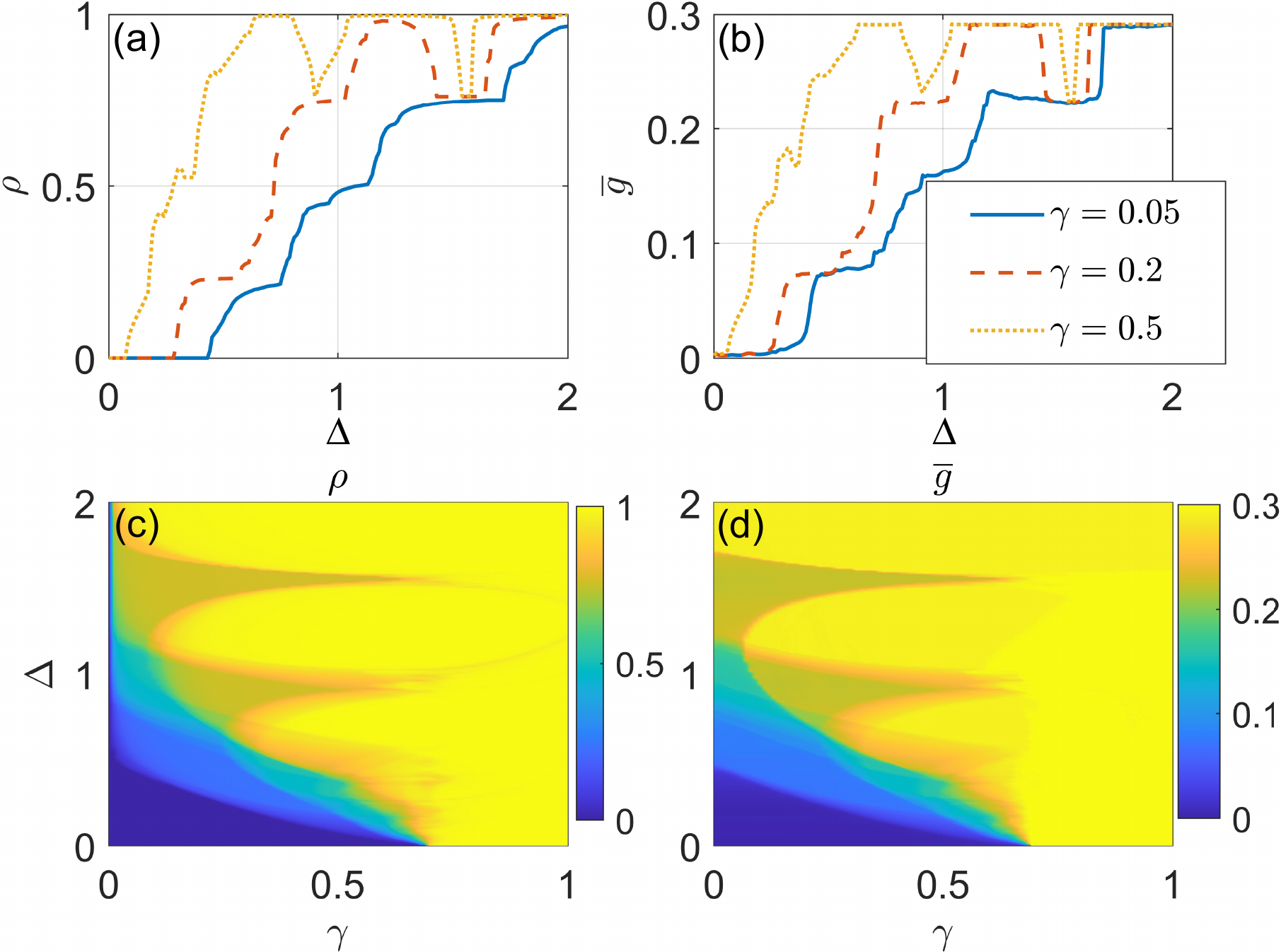}
		\par\end{centering}
	\caption{Spectral properties and AGRs of M3, obtained for
		a lattice of length $L=2584$ and under the PBC. Other system parameters
		are $(J,V)=(1,1)$. The solid, dashed and dotted lines in (a) and (b)
		show the DOSs {[}Eq.~(\ref{eq:DOS}){]} and AGRs {[}Eq.~(\ref{eq:AveAGR}){]} versus
		the onsite dimerization $\Delta$ for different $\gamma$.
		In (c) and (d), the DOSs and AGRs versus both $\gamma$ and $\Delta$ show consistent patterns for $\gamma\neq0$.\label{fig:M3EDOS}}
\end{figure}

In the last example, we consider the case in which 
the non-Hermiticity is introduced by an imaginary phase shift
in the AAH model with a staggered onsite potential,
leading to the M3 in Table \ref{tab:Mod}. The Hamiltonian
of M3 takes the form $H=\sum_{n}(|n\rangle\langle n+1|+{\rm H.c.}+V_{n}|n\rangle\langle n|)$ with $V_{n}=V\cos(2\pi\alpha n+i\gamma)+(-1)^{n}\Delta$.
The eigenvalue equation reads $\psi_{n+1}+\psi_{n-1}+V_{n}\psi_{n}=E\psi_{n}$.
Since $V_{n}=V_{-n}^{*}$, $H$ here also possesses
the ${\cal PT}$-symmetry and its spectrum could be real. With the
increase of the staggering strength $\Delta$, spectral and
localization transitions would also occur in M3 as demonstrated
below. Most notably, triggered by onsite dimerization, the system 
is found to be able to roam alternately between localized 
and critical mobility edge phases in a non-monotonic way.

\begin{figure}
	\begin{centering}
		\includegraphics[scale=0.48]{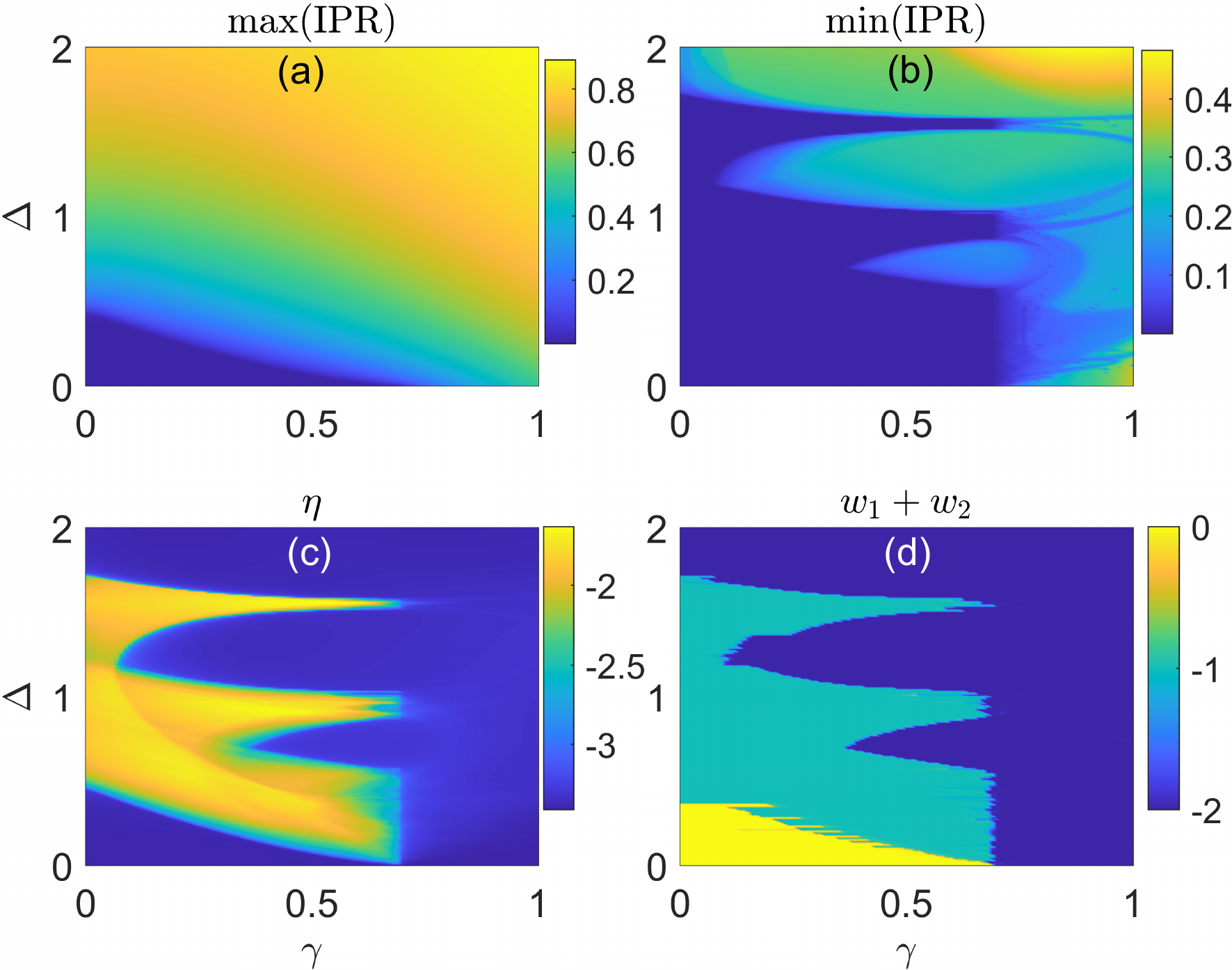}
		\par\end{centering}
	\caption{State properties and winding numbers of M3 obtained under the PBC. 
		System parameters are $(J,V)=(1,1)$. (a)--(c) show the maximum
		{[}Eq.~(\ref{eq:IPRmax}){]}, minimum {[}Eq.~(\ref{eq:IPRmin}){]}
		of IPRs and $\eta$ {[}Eq.~(\ref{eq:ETA}){]} at different system parameters $(\gamma,\Delta)$,
		computed with a lattice size $L=2584$.
		(d) shows the topological phase diagram,
		in which yellow, green and blue regions correspond to the extended,
		critical and localized NHQC phases with spectral winding numbers $(w_{1},w_{2})=(0,0)$,
		$(-1,0)$, and $(-1,-1)$, respectively, computed with a lattice size $L=610$.\label{fig:M3WN}}
\end{figure}

In Fig.~\ref{fig:M3E}, we present selected spectra of M3 under different strengths of lattice
dimerization. We observe that with a finite amount of imaginary phase shift $i\gamma$,
the number of real eigenvalues in the spectrum could change with $\Delta$ in a highly non-monotonic manner.
Specifically, the real eigenvalues in the spectrum vanish when $\Delta$ changes from $0.2$ to $0.7$, but re-emerge
when $\Delta$ goes from $0.7$ to $0.9$. This process could repeat until $\Delta$ is large enough and
all states end in taking complex energies. These rich and non-monotonic behaviors of spectrum are closely
related to the reentrant localization and topological transitions of M3.

In Fig.~\ref{fig:M3EDOS}, we present the DOSs with nonzero imaginary parts of energy $\rho$
and AGRs $\overline{g}$ versus the phase shift $\gamma$ and onsite dimerization $\Delta$.
Close to the Hermitian limit ($\gamma\simeq0$), we find a transition of the system from real ($\rho=0$) to complex ($\rho>0$) spectrum with the increase of $\Delta$, which goes together with the change of $\overline{g}$ from zero to a finite value $0<\overline{g}\lesssim0.3$. The further increase of $\Delta$ causes a second transition, after which $\rho$ and $\overline{g}$ approach $1$ and $0.3$. Surprisingly, when $\gamma$ is away from zero but still smaller than the critical value $\gamma_c=\ln|2J/V|$, we find $\rho$ and $\overline{g}$ to show reentrant behaviors between the phases with $\rho\in(0,1)$, $\overline{g}\in(0,0.3)$, and $\rho=1$, $\overline{g}\simeq0.3$, which are clearly demonstrated by the dashed and dotted lines shown in Figs.~\ref{fig:M3EDOS}(a)--\ref{fig:M3EDOS}(b). These reentrant transitions, which are absent if $\Delta=0$ in M3, are expected to be consistent with the alternation of the system between localized and critical phases through a series of topological localization transitions, as confirmed by the results presented in Figs.~\ref{fig:M3WN}(a)--\ref{fig:M3WN}(d). For example, near $\gamma=0.5$, the system is found to follow a sequence of transitions from extended $\rightarrow$ critical $\rightarrow$ localized $\rightarrow$ critical $\rightarrow$ localized $\rightarrow$ critical, and finally ends in a localized phase with the increase of onsite dimerization $\Delta$. This transition sequence is further accompanied by the changes of winding numbers following $(w_1,w_2)=(0,0)$ $\rightarrow$ $(-1,0)$ $\rightarrow$ $(-1,-1)$ $\rightarrow$ $(-1,0)$ $\rightarrow$ $(-1,-1)$ $\rightarrow$ $(-1,0)$, and finally arrives at $(w_1,w_2)=(-1,-1)$.
Note that for M3, the phase shift in Eq.~(\ref{eq:W})
is introduced by setting $i\gamma\rightarrow i\gamma+2\theta/L$ in $V_n$.
Therefore, we conclude that the lattice dimerization could not only produce critical phases with mobility edges, but also induce multiple and reentrant topological localization transitions in NHQCs. Notably, the reentrant transitions found here are different from those observed in Hermitian quasicrystals~\cite{DimerQC2}, as in our case these transitions are only found at finite amounts of imaginary phase shift $i\gamma$, implying that they are unique phenomena originated from the interplay between non-Hermiticity and spatial dimerization. Moreover, the alternating jumps of topological order parameters endow richer structures to the reentrant transitions in NHQCs compared with their Hermitian counterparts~\cite{DimerQC1,DimerQC2,DimerQC3}, providing us with more insights to non-Hermitian disordered systems from a topological perspective.

\section{Summary\label{sec:Sum}}

In this work, we found lattice dimerization induced critical phases
with mobility edges and multiple reentrant localization transitions
in NHQCs. These findings were explicitly
demonstrated in non-Hermitian extensions of the AAH model with staggered
onsite potentials or dimerized hopping amplitudes. Especially, the
interplay between dimerization and non-Hermitian effects works as a
flexible knob to control the phase transitions and critical properties
of NHQCs. Moreover, topological order parameters
were employed to precisely distinguish extended, critical, localized
phases of NHQCs and capture the transitions
among them. Put together, our results unveil the richness of NHQC
phases in the presence of lattice dimerization, uncover the
chance of generating reentrant localization transtions in NHQCs, and
suggest a way to manipulate the transport nature of NHQCs
by tuning dimerization effects.
In Refs.~\cite{DimerQC1,DimerQC2,DimerQC3}, multiple and/or reentrant localization transitions
are found in quasicrystals with lattice dimerization or during the process in which one type of quasicrystal changes
to another. In all the cases, the systems under consideration are described by Hermitian Hamiltonians.
Compared with these former studies,
our work uncovers the role played by the collaboration of non-Hermitian effects and lattice dimerization in generating reentrant
localization and topological phase transitions that are unique to non-Hermitian settings.
In future work,
it would be interesting to consider lattice dimerization in different
types of NHQC models, such as those with nonreciprocal
hoppings, and explore more exotic phases of dimerized NHQCs
in the presence of many-body interactions and time-periodic drives.
The possible interplay between lattice dimerization and the recently found
pseudo mobility edges~\cite{ZhouPME} due to non-Hermitian skin effects also deserve more
thorough explorations.

\begin{acknowledgments}
L.Z. is supported by the National Natural Science Foundation of China (Grant No.~11905211), the Young Talents Project at Ocean University of China (Grant No.~861801013196), and the Applied Research Project of Postdoctoral Fellows in Qingdao (Grant No.~861905040009).
\end{acknowledgments}
%



\begin{thebibliography}{99}
	
	\bibitem{NHRev1} E. J. Bergholtz, J. C. Budich, and F. K. Kunst,
	Exceptional topology of non-Hermitian systems, Rev. Mod. Phys. \textbf{93},
	015005 (2021).
	
	\bibitem{NHRev2} Y. Ashida, Z. Gong, and M. Ueda, Non-Hermitian physics,
	Adv. Phys. \textbf{69}, 249-435 (2020).
	
	\bibitem{NHRev3} R. El-Ganainy, K. G. Makris, M. Khajavikhan, Z.
	H. Musslimani, S. Rotter, and D. N. Christodoulides, Non-Hermitian
	physics and PT symmetry, Nat. Phys. \textbf{14}, 11-19 (2018).
	
	\bibitem{NHRev4} A. Ghatak and T. Das, New topological invariants
	in non-Hermitian systems, J. Phys.: Condens. Matter \textbf{31}, 263001
	(2019).
	
	\bibitem{NHRev5} V. M. Martinez Alvarez, J. E. Barrios Vargas, M.
	Berdakin, and L. E. F. Foa Torres, Topological states of non-Hermitian
	systems, Eur. Phys. J. Spec. Top. \textbf{227}, 1295 (2018).
	
	\bibitem{NHRev6} C. Coulais, R. Fleury, and J. van
	Wezel, Topology and broken Hermiticity, Nat. Phys. \textbf{17}, 9-13
	(2021).
	
	\bibitem{EP1} M. V. Berry, Physics of Nonhermitian Degeneracies, Czech. J. Phys. {\bf 54}, 1039 (2004).
	
	\bibitem{EP2} W. D. Heiss, The physics of exceptional points, J. Phys. A: Math. Theor. {\bf 45}, 444016 (2012).
	
	\bibitem{EP3} M.-A. Miri and A. Al\`u, Exceptional points in optics	and photonics, Science {\bf 363}, eaar7709 (2019).
	
	\bibitem{NHSE1} S. Yao and Z. Wang, Edge States and Topological Invariants
	of Non-Hermitian Systems, Phys. Rev. Lett. \textbf{121}, 086803 (2018).
	
	\bibitem{NHSE2} F. K. Kunst, E. Edvardsson, J. C. Budich, and E.
	J. Bergholtz, Biorthogonal Bulk-Boundary Correspondence in Non-Hermitian
	Systems, Phys. Rev. Lett. \textbf{121}, 026808 (2018).
	
	\bibitem{NHSE3} V. M. Martinez Alvarez, J. E. Barrios Vargas, and
	L. E. F. Foa Torres, Non-Hermitian robust edge states in one dimension:
	anomalous localization and eigenspace condensation at exceptional
	points, Phys. Rev. B \textbf{97}, 121401(R) (2018).
	
	\bibitem{NHSE4} C. H. Lee and R. Thomale, Anatomy of skin modes and
	topology in non-Hermitian systems, Phys. Rev. B \textbf{99}, 201103(R)
	(2019).
	
	\bibitem{HNM1} N. Hatano and D. R. Nelson, Localization Transitions in Non-Hermitian Quantum Mechanics,
	Phys. Rev. Lett. {\bf 77}, 570 (1996).
	
	\bibitem{HNM2} J. Feinberg and A. Zee, Non-Hermitian localization and delocalization,
	Phys. Rev. E {\bf 59}, 6433 (1999).
	
	\bibitem{HNM3} J. Feinberg and A. Zee, Spectral curves of non-hermitian hamiltonians,
	Nucl. Phys. B {\bf 552}, 599-623 (1999).
	
	\bibitem{HNM4} N. Hatano and J. Feinberg, Chebyshev-polynomial expansion of the localization length of Hermitian and non-Hermitian random chains,
	Phys. Rev. E {\bf 94}, 063305 (2016).
	
	\bibitem{NHClass1} Z. Gong, Y. Ashida, K. Kawabata, K. Takasan, S.
	Higashikawa, and M. Ueda, Topological Phases of Non-Hermitian Systems,
	Phys. Rev. X \textbf{8}, 031079 (2018).
	
	\bibitem{NHClass2} K. Kawabata, K. Shiozaki, M. Ueda and M. Sato,
	Symmetry and Topology in Non-Hermitian Physics, Phys. Rev. X \textbf{9},
	041015 (2019).
	
	\bibitem{NHClass3} H. Zhou and J. Y. Lee, Periodic table for topological
	bands with non-Hermitian symmetries, Phys. Rev. B \textbf{99}, 235112
	(2019).
	
	\bibitem{NHClass4} C. C. Wojcik, X. Sun, T. Bzdu$\check{\rm s}$ek, and S. Fan,
	Homotopy characterization of non-Hermitian Hamiltonians, Phys. Rev.
	B \textbf{101}, 205417 (2020).
	
	\bibitem{NHClass5} K. Shiozaki and S. Ono, Symmetry indicator in
	non-Hermitian systems, Phys. Rev. B \textbf{104}, 035424 (2021).
	
	\bibitem{NHExp1} W. Gou, T. Chen, D. Xie, T. Xiao, T.-S. Deng, B.
	Gadway, W. Yi, and B. Yan, Tunable Nonreciprocal Quantum Transport
	through a Dissipative Aharonov-Bohm Ring in Ultracold Atoms, Phys.
	Rev. Lett. \textbf{124}, 070402 (2020).
	
	\bibitem{NHExp2} J. Li, A. K. Harter, J. Liu, L. d. Melo, Y. N. Joglekar,
	and L. Luo, Observation of parity-time symmetry breaking transitions
	in a dissipative Floquet system of ultracold atoms, Nat. Commun. \textbf{10},
	855 (2019).
	
	\bibitem{NHExp3} Y. Xu, S.-T. Wang, and L.-M. Duan, Weyl Exceptional
	Rings in a Three-Dimensional Dissipative Cold Atomic Gas, Phys. Rev.
	Lett. \textbf{118}, 045701 (2017).
	
	\bibitem{NHExp4} J. M. Zeuner, M. C. Rechtsman, Y. Plotnik, Y. Lumer,
	S. Nolte, M. S. Rudner, M. Segev, and A. Szameit, Observation of a
	Topological Transition in the Bulk of a Non-Hermitian System. Phys.
	Rev. Lett. \textbf{115}, 040402 (2015).
	
	\bibitem{NHExp5} S. Weimann, M. Kremer, Y. Plotnik, Y. Lumer, S.
	Nolte, K. G. Makris, M. Segev, M. C. Rechtsman, and A. Szameit, Topologically
	protected bound states in photonic parity-time-symmetric crystals,
	Nat. Mater. \textbf{16}, 433-438 (2017).
	
	\bibitem{NHExp6} K. Wang, X. Qiu, L. Xiao, X. Zhan, Z. Bian, B. C.
	Sanders, W. Yi, and P. Xue, Observation of emergent momentum-time
	skyrmions in parity-time-symmetric non-unitary quench dynamics, Nat.
	Commun. \textbf{10}, 2293 (2019).
	
	\bibitem{NHExp7} L. Xiao, T. Deng, K. Wang, G. Zhu, Z. Wang, W. Yi,
	and P. Xue, Non-Hermitian bulk-boundary correspondence in quantum
	dynamics, Nat. Phys. \textbf{16}, 761-766 (2020).
	\bibitem{NHExp11} W. Zhu, X. Fang, D. Li, Y. Sun, Y. Li, Y. Jing,
	and H. Chen, Simultaneous Observation of a Topological Edge State
	and Exceptional Point in an Open and Non-Hermitian Acoustic System,
	Phys. Rev. Lett. \textbf{121}, 124501 (2018).
	
	\bibitem{NHExp12} C. Shen, J. Li, X. Peng, and S. A. Cummer, Synthetic
	exceptional points and unidirectional zero reflection in non-Hermitian
	acoustic systems, Phys. Rev. Mater. \textbf{2}, 125203 (2018).
	
	\bibitem{NHExp13} H. Gao, H. Xue, Q. Wang, Z. Gu, T. Liu, J. Zhu,
	and B. Zhang, Observation of topological edge states induced solely
	by non-Hermiticity in an acoustic crystal, Phys. Rev. B \textbf{101},
	180303(R) (2020).
	
	\bibitem{NHExp8} T. Hofmann, T. Helbig, F. Schindler, N. Salgo, M.
	Brzezi\'{n}ska, M. Greiter, T. Kiessling, D. Wolf, A. Vollhardt, A.
	Kaba$\check{\rm s}$i, C. H. Lee, A. Bilu$\check{\rm s}$i\'{c}, R. Thomale, and T. Neupert, Reciprocal
	skin effect and its realization in a topolectrical circuit, Phys.
	Rev. Res. \textbf{2}, 023265 (2020).
	
	\bibitem{NHExp9} T. Helbig, T. Hofmann, S. Imhof, M. Abdelghany,
	T. Kiessling, L. W. Molenkamp, C. H. Lee, A. Szameit, M. Greiter,
	and R. Thomale, Generalized bulk-boundary correspondence in non-Hermitian
	topolectrical circuits, Nat. Phys. \textbf{16}, 747-750 (2020).
	
	\bibitem{NHExp10} S. Liu, S. Ma, C. Yang, L. Zhang, W. Gao, Y. J.
	Xiang, T. J. Cui, and S. Zhang, Gain- and Loss-Induced Topological
	Insulating Phase in a Non-Hermitian Electrical Circuit, Phys. Rev.
	Appl. \textbf{13}, 014047 (2020).
	
	\bibitem{NHExp14} Y. Wu, W. Liu, J. Geng, X. Song, X. Ye, C.-K. Duan,
	X. Rong, and J. Du, Observation of parity-time symmetry breaking in
	a single-spin system, Science \textbf{364}, 878-880 (2019).
	
	\bibitem{TILZ1} G. Harari, M. A. Bandres, Y. Lumer, M. C. Rechtsman,
	Y. D. Chong, M. Khajavikhan, D. N. Christodoulides, and M. Segev,
	Topological insulator laser: Theory. Science \textbf{359}, 4003 (2018).
	
	\bibitem{TILZ2} M. A. Bandres, S. Wittek, G. Harari, M. Parto, J.
	Ren, M. Segev, D. N. Christodoulides, and M. Khajavikhan, Topological
	insulator laser: Experiments. Science \textbf{359}, 4005 (2018).
	
	\bibitem{TILZ3} Y. V. Kartashov and D. V. Skryabin, Two-Dimensional
	Topological Polariton Laser, Phys. Rev. Lett. \textbf{122}, 083902
	(2019).
	
	\bibitem{NHSens1} J. Wiersig, Enhancing the Sensitivity of Frequency
	and Energy Splitting Detection by Using Exceptional Points: Application
	to Microcavity Sensors for Single-Particle Detection, Phys. Rev. Lett.
	\textbf{112}, 203901 (2014).
	
	\bibitem{NHSens2} H.-K. Lau and A. A. Clerk, Fundamental limits and
	non-reciprocal approaches in non-Hermitian quantum sensing, Nat. Commun.
	\textbf{9}, 4320 (2018).
	
	\bibitem{NHSens3} H. Hodaei, A. U. Hassan, S. Wittek, H. Garcia-Gracia,
	R. El-Ganainy, D. N. Christodoulides, and M. Khajavikhan, Enhanced
	sensitivity at higher-order exceptional points, Nature \textbf{548},
	187-191 (2017).
	
	\bibitem{NHSens4} W. Chen, S. K. \"Ozdemir, G. Zhao, J. Wiersig, and
	L. Yang, Exceptional points enhance sensing in an optical microcavity,
	Nature \textbf{548}, 192-196 (2017).
	
	\bibitem{NHQC0} P. Sarnak, Spectral Behavior of Quasi
		Periodic Potentials, Commun. Math. Phys. {\bf 84}, 377-401 (1982).
	
	\bibitem{NHQC12} A. Jazaeri and I. I. Satija, Localization transition
	in incommensurate non-Hermitian systems, Phys. Rev. E \textbf{63},
	036222 (2001).
	
	\bibitem{NHQC10} Q. Zeng, S. Chen, and R. L\"u, Anderson localization
	in the non-Hermitian Aubry-Andr\'e-Harper model with physical gain and
	loss, Phys. Rev. A \textbf{95}, 062118 (2017).
	
	\bibitem{NHQC5} H. Jiang, L. Lang, C. Yang, S. Zhu, and S. Chen,
	Interplay of non-Hermitian skin effects and Anderson localization
	in nonreciprocal quasiperiodic lattices, Phys. Rev. B \textbf{100},
	054301 (2019).
	
	\bibitem{NHQC1} S. Longhi, Topological Phase Transition in non-Hermitian
	Quasicrystals, Phys. Rev. Lett. \textbf{122}, 237601 (2019).
	
	\bibitem{NHQC2} S. Longhi, Metal-insulator phase transition in a
	non-Hermitian Aubry-Andr\'e-Harper model, Phys. Rev. B \textbf{100},
	125157 (2019).
	
	\bibitem{NHQC3} T. Liu, H. Guo, Y. Pu, and S. Longhi, Generalized
	Aubry-Andr\'e self-duality and mobility edges in non-Hermitian quasiperiodic
	lattices, Phys. Rev. B \textbf{102}, 024205 (2020).
	
	\bibitem{NHQC6} Y. Liu, X.-P. Jiang, J. Cao, and S. Chen, Non-Hermitian
	mobility edges in one-dimensional quasicrystals with parity-time symmetry,
	Phys. Rev. B \textbf{101}, 174205 (2020).
	
	\bibitem{NHQC13} Q. Zeng, Y. Yang, and R. L\"u, Topological phases
	in one-dimensional nonreciprocal superlattices, Phys. Rev. B \textbf{101},
	125418 (2020).
	
	\bibitem{NHQC14} Q. Zeng, Y. Yang, and Y. Xu, Topological phases
	in non-Hermitian Aubry-Andr\'e-Harper models, Phys. Rev. B \textbf{101},
	020201(R) (2020).
	
	\bibitem{NHQC18} L. Zhai, S. Yin, and G. Huang, Many-body localization
	in a non-Hermitian quasiperiodic system, Phys. Rev. B \textbf{102},
	064206 (2020).
	
	\bibitem{NHQC19} Q. Zeng and Y. Xu, Winding numbers and generalized
	mobility edges in non-Hermitian systems, Phys. Rev. Res. \textbf{2},
	033052 (2020).
	
	\bibitem{NHQC4} S. Longhi, Phase transitions in a non-Hermitian Aubry-Andr\'e-Harper
	model, Phys. Rev. B \textbf{103}, 054203 (2021).
	
	\bibitem{NHQC7} Y. Liu, Y. Wang, Z. Zheng, and S. Chen, Exact non-Hermitian
	mobility edges in one-dimensional quasicrystal lattice with exponentially
	decaying hopping and its dual lattice, Phys. Rev. B \textbf{103},
	134208 (2021).
	
	\bibitem{NHQC8} Y. Liu, Y. Wang, X. Liu, Q. Zhou, and S. Chen, Exact
	mobility edges, PT-symmetry breaking, and skin effect in one-dimensional
	non-Hermitian quasicrystals, Phys. Rev. B \textbf{103}, 014203 (2021).
	
	\bibitem{NHQC9} Z. Xu and S. Chen, Dynamical evolution in a one-dimensional
	incommensurate lattice with PT symmetry, Phys. Rev. A\textbf{ 103},
	043325 (2021).
	
	\bibitem{NHQC11} X. Cai, Boundary-dependent self-dualities, winding
	numbers, and asymmetrical localization in non-Hermitian aperiodic
	one-dimensional models, Phys. Rev. B \textbf{103}, 014201 (2021).
	
	\bibitem{NHQC15} L. Tang, G. Zhang, L. Zhang, and D. Zhang, Localization
	and topological transitions in non-Hermitian quasiperiodic lattices,
	Phys. Rev. A \textbf{103}, 033325 (2021).
	
	\bibitem{NHQC16} T. Liu, S. Cheng, H. Guo, and X. Gao, Fate of Majorana
	zero modes, exact location of critical states, and unconventional
	real-complex transition in non-Hermitian quasiperiodic lattices, Phys.
	Rev. B \textbf{103}, 104203 (2021).
	
	\bibitem{NHQC17} L. Zhai, G. Huang, and S. Yin, Cascade of the delocalization
	transition in a non-Hermitian interpolating Aubry-Andr\'e-Fibonacci
	chain, Phys. Rev. B \textbf{104}, 014202 (2021).
	
	\bibitem{NHQC20} L. Zhou and W. Han, Non-Hermitian quasicrystal in dimerized lattices,
	Chin. Phys. B {\bf 30}, 100308 (2021).
	
	\bibitem{NHQC201} Z.-H. Wang, F. Xu, L. Li, D. Xu, and B. Wang, Unconventional real-complex spectral transition and Majorana zero modes
	in nonreciprocal quasicrystals, Phys. Rev. B {\bf 104}, 174501 (2021).
	
	\bibitem{NHQC21} Y. Liu, Q. Zhou, and S. Chen, Localization transition,
	spectrum structure, and winding numbers for one-dimensional non-Hermitian
	quasicrystals, Phys. Rev. B \textbf{104}, 024201 (2021).
	
	\bibitem{NHQC22} X. Cai, Localization and topological phase transitions
	in non-Hermitian Aubry-Andr\'e-Harper models with \emph{p}-wave pairing,
	Phys. Rev. B \textbf{103}, 214202 (2021).
	
	\bibitem{NHQC23} S. Longhi, Non-Hermitian Maryland model, Phys. Rev.
	B \textbf{103}, 224206 (2021).
	
	\bibitem{NHQC24} L. Zhou, Floquet engineering of topological localization transitions and mobility edges in one-dimensional non-Hermitian quasicrystals,
	Phys. Rev. Research {\bf 3}, 033184 (2021).
	
	\bibitem{NHQC26} A. P. Acharya, A. Chakrabarty, and D. K. Sahu, Localization,
	PT-Symmetry Breaking and Topological Transitions in non-Hermitian
	Quasicrystals, Phys. Rev. B {\bf 105}, 014202 (2022).
	
	\bibitem{NHQC27} L. Zhou and J. Gu, Topological delocalization transitions and mobility edges in the nonreciprocal Maryland model, J. Phys.: Condens. Matter {\bf 34}, 115402 (2022).
	
	\bibitem{NHQC25} X. Xia, K. Huang, S. Wang, and X. Li, Exact mobility edges in the non-Hermitian $t_1-t_2$ model: Theory and possible experimental realizations, Phys. Rev. B {\bf 105}, 014207 (2022).
	
	\bibitem{LRQC1} R. B. Diener, G. A. Georgakis, J. Zhong, M. Raizen, and Q. Niu, Transition between extended and localized states in a one dimensional incommensurate optical lattice, Phys. Rev. A {\bf 64}, 033416 (2001).
	
	\bibitem{LRQC2} D. J. Boers, B. Goedeke, D. Hinrichs, and M. Holthaus, Mobility edges in bichromatic optical lattices, Phys. Rev. A {\bf 75},
	063404 (2007).
	
	\bibitem{LRQC3} X. Li, X. Li, and S. Das Sarma, Mobility edges in one dimensional bichromatic incommensurate potentials, Phys.
	Rev. B {\bf 96}, 085119 (2017).
	
	\bibitem{DimerQC1} V. Goblot, A. \v{S}trkalj, N. Pernet, J. L. Lado, C. Dorow, A. Lema\^{i}tre, L. Le Gratiet, A. Harouri, I. Sagnes, S. Ravets, A. Amo, J. Bloch, and O. Zilberberg, Emergence of criticality through a cascade of delocalization transitions in quasiperiodic chains, Nat. Phys. {\bf 16}, 832 (2020).
	
	\bibitem{DimerQC2} S. Roy, T. Mishra, B. Tanatar, and S. Basu, Reentrant Localization Transition in a Quasiperiodic Chain, Phys. Rev. Lett. {\bf 126}, 106803 (2021).
	
	\bibitem{DimerQC3} A. Padhan, M. K. Giri, S. Mondal and T. Mishra, Emergence of multiple localization transitions in a one-dimensional quasiperiodic
	lattice, arXiv:2109.09621.
	
	\bibitem{NHRMT1} I. Y. Goldsheid and B. A. Khoruzhenko, Distribution
	of Eigenvalues in Non-Hermitian Anderson Models, Phys. Rev. Lett.
	\textbf{80}, 2897 (1998).
	
	\bibitem{NHRMT2} L. G. Molinari, Non-Hermitian spectra and Anderson
	localization, J. Phys. A: Math. Theor. \textbf{42}, 265204 (2009).
	
	\bibitem{NHRMT3} H. Markum, R. Pullirsch, and T. Wettig, Non-Hermitian
	Random Matrix Theory and Lattice QCD with Chemical Potential, Phys.
	Rev. Lett. \textbf{83}, 484 (1999).
	
	\bibitem{NHRMT4} J. T. Chalker and B. Mehlig, Eigenvector Statistics
	in Non-Hermitian Random Matrix Ensembles, Phys. Rev. Lett. \textbf{81},
	3367 (1998).
	
	\bibitem{NHRMT5} R. Hamazaki, K. Kawabata, N. Kura, and M. Ueda,
	Universality classes of non-Hermitian random matrices, Phys. Rev.
	Res. \textbf{2}, 023286 (2020).
	
	\bibitem{CMTBook1} S. M. Girvin and K. Yang, Modern Condensed Matter
	Physics (Cambridge University Press, New York, 2019).
	
	\bibitem{AAH1} S. Aubry and G. Andr\'e, Analyticity breaking and Anderson
	localization in incommensurate lattices, Ann. Israel Phys. Soc.
	{\bf 3}, 133 (1980).
	
	\bibitem{AAH2} P. G. Harper, Single band motion of conduction electrons in
	a uniform magnetic field, Proc. Phys. Soc. London A {\bf 68}, 874 (1955).
	
	\bibitem{AAH3} J. B. Sokoloff, Unusual band structure, wave function and
	electrical conductance in crystals with incommensurate periodic potentials, Phys. Rep. {\bf 126}, 189 (1985).
	
		\bibitem{SSH1} M. Atala, M. Aidelsburger, J. T. Barreiro, D. Abanin, T. Kitagawa, 
		E. Demler, and I. Bloch, Direct measurement of the Zak phase in topological 
		Bloch bands Nat. Phys. {\bf 9}, 795-800 (2013).
	
	\bibitem{SSH2} S. Nakajima, T. Tomita, S. Taie, T. Ichinose, H. Ozawa, L. Wang,
		M. Troyer, and Y. Takahashi, Topological Thouless pumping of ultracold fermions,
		Nat. Phys. {\bf 12}, 296-300 (2016).
	
	\bibitem{SSH3} M. Lohse, C. Schweizer, O. Zilberberg, M. Aidelsburger, and I. Bloch,
		A Thouless quantum pump with ultracold bosonic atoms in an optical superlattice,
		Nat. Phys. {\bf 12}, 350-354 (2016).
	
	\bibitem{ZhouPME} S. Mu, L. Zhou, L. Li, and J. Gong, Non-Hermitian pseudo mobility edge in a coupled chain system, arXiv:2111.11914.
	
	
\end{thebibliography}
\end{document}